# Dilaton field and cosmic wave propagation

Wei-Tou Ni

*Center for Gravitation and Cosmology, Department of Physics,*
*National Tsing Hua University, Hsinchu, Taiwan, 30013, Republic of China*

**Abstract**

We study the electromagnetic wave propagation in the joint dilaton field and axion field. Dilaton field induces amplification/attenuation in the propagation while axion field induces polarization rotation. The amplification/attenuation induced by dilaton is independent of the frequency (energy) and the polarization of electromagnetic waves (photons). From observations, the agreement with and the precise calibration of the cosmic microwave background (CMB) to blackbody radiation constrains the fractional change of dilaton $|\Delta\psi|/\psi$ to less than about $8 \times 10^{-4}$ since the time of the last scattering surface of the CMB.

# 1. Introduction

The concept of metric in Minkowski's spacetime formulation [1] of special relativity together with the proposal of Einstein Equivalence Principle [2] facilitated the genesis of general relativity. While metric field is the basic dynamical entity of gravity in general relativity, Einstein Equivalence Principle (EEP) dictates the coupling of gravity to matter. In putting Maxwell equations into a form compatible with general relativity, Einstein noticed that the equations can be formulated in a form independent of the metric gravitational potential in 1916 [3, 4] shortly after his completion of general relativity with further developments worked out by Weyl [5], Murnaghan [6], Kottler [7] and Cartan [8]. In this introduction, we first review the premetric formulation of electromagnetism and then address to the issue of construction of the core metric, the dilaton field and the axion field from the nonbirefringent wave propagation.

*1.1. Premetric formulation of electromagnetism*

Maxwell equations in terms of field strength $F_{kl}$ (***E***, ***B***) and excitation (density) $H^{ij}$ (***D***, ***H***) do not need metric as primitive concept. Field strength $F_{kl}$ (***E***, ***B***) and excitation $H^{ij}$ (***D***, ***H***) can all be independently operational defined (See, e. g., Hehl and Obukhov [9]). Maxwell equations of macroscopic/spacetime electrodynamics expressed in terms of these quantities are

$$H^{ij}{}_{,j} = -4\pi J^i, \tag{1a}$$
$$e^{ijkl} F_{jk,l} = 0, \tag{1b}$$

where $J^k$ is the charge 4-current density and $e^{ijkl}$ the completely anti-symmetric tensor density with $e^{0123} = 1$ (See, e. g., Hehl and Obukhov [9]). We use units with the light velocity $c$ equal to 1. To complete this set of equations, one needs the constitutive relation between the excitation and the field:

$$H^{ij} = (1/2) \chi^{ijkl} F_{kl}. \tag{2}$$

Both $H^{ij}$ and $F_{kl}$ are antisymmetric, hence $\chi^{ijkl}$ must be antisymmetric in $i$ and $j$, and $k$ and $l$. Therefore the constitutive tensor density $\chi^{ijkl}$ has 36 independent components. A general linear constitutive tensor density $\chi^{ijkl}$ in electrodynamics can be decomposed into principal part (P), axion part (Ax) and skewon part (Sk) [9]:

$$\chi^{ijkl} = {}^{(P)}\chi^{ijkl} + {}^{(Sk)}\chi^{ijkl} + {}^{(Ax)}\chi^{ijkl}, \qquad (\chi^{ijkl} = -\chi^{jikl} = -\chi^{ijlk}) \qquad (3)$$

with

$$^{(P)}\chi^{ijkl} = (1/6)[2(\chi^{ijkl} + \chi^{klij}) - (\chi^{iklj} + \chi^{ljik}) - (\chi^{iljk} + \chi^{jkil})], \qquad (4a)$$

$$^{(Ax)}\chi^{ijkl} = \chi^{[ijkl]} = \varphi\, e^{ijkl}, \qquad (4b)$$

$$^{(Sk)}\chi^{ijkl} = (1/2)(\chi^{ijkl} - \chi^{klij}). \qquad (4c)$$

Decomposition (3) is unique. The principal part has 20 degrees of freedom. The axion part has one degree of freedom. The skewon part has 15 degrees of freedom. The skewon field was proposed by Hehl, Obukhov and Rubilar [10, 11] and has been studied extensively [12-19]. In a recent paper [17], we have studied the electromagnetic wave propagation and the observational constraint on the skewon field in the weak field limit. We found that the CMB spectrum measurement gives very stringent constraint on the Type I skewons. From the dispersion relation we show that no dissipation/no amplification condition implies that the additional skewon field must be of Type II. For Type I skewon field, the dissipation/amplification is proportional to the frequency and the CMB spectrum would deviate from Planck spectrum in shape. From the high precision agreement of the CMB spectrum to 2.755 K Planck spectrum, we constrain the Type I cosmic skewon field $|^{(SkI)}\chi^{ijkl}|$ to ≤ a few × $10^{-35}$.

In the skewonless case, the well-observed nonbirefringence condition in the astrophysical/cosmological propagation constrains the spacetime constitutive tensor density $\chi^{ijkl}$ to a core metric plus dilaton field and axion field very stringently.

The axion part of the constitutive tensor gives pseudoscalar-photon interaction [20]. This pseudoscalar-photon interaction has been studied in detail in [21, 22]. It induces CPR (Cosmic Polarization Rotation). The astrophysical and cosmological constraints on CPR and, hence, on axion field are reviewed in [23-26]. In macroscopic electrodynamics, Hehl, Obukhov, Rivera and Schmid [27] have studied and clarified that the chromium sesquioxide $Cr_2O_3$ crystal is an axionic medium. Their paper has explicitly demonstrated that the four-dimensional pseudoscalar $\varphi$ (4b) exists in macroscopic electrodynamics. In view of the existence of axionic material [27], searching for dilatonic material and skewonic material in macroscopic electrodynamics would also be interesting.

*1.2. Derivation of spacetime structure from premetric electrodynamics*

The issue here is that how to (with what conditions can we) reach a metric or,

owing to conformal invariance, how to reach a Riemannian light cone (a core metric up to conformal invariance) from the constitutive tensor. This issue has been studied rather thoroughly in the skewonless case, i.e. in the case $\chi^{ijkl}$ is symmetric under the exchange of the index pairs *ij* and *kl*. In this case, the Maxwell equations can be derived from the Lagrangian density $L$ (= $L_I^{(EM)}$ + $L_I^{(EM-P)}$) with the electromagnetic field Lagrangian density $L_I^{(EM)}$ and the field-current interaction Lagrangian density $L_I^{(EM-P)}$ given by

$$L_I^{(EM)} = -(1/(8\pi))H^{ij} F_{ij} = -(1/(16\pi))\chi^{ijkl} F_{ij} F_{kl}, \qquad (5)$$
$$L_I^{(EM-P)} = -A_k J^k, \qquad (6)$$

where $\chi^{ijkl} = \chi^{klij} = -\chi^{jikl}$ is a tensor density of the gravitational fields or matter fields to be investigated, $A_i$ the electromagnetic 4-potential, $F_{ij} \equiv A_{j,i} - A_{i,j}$ the electromagnetic field strength tensor and comma denoting partial derivation [20, 28]. We note that only the part of $\chi^{ijkl}$ which is symmetric under the interchange of index pairs *ij* and *kl* contributes to the Lagrangian density, and hence, there is no skewon contribution.

One way to reach a core metric for spacetime constitutive tensor is through Galileo weak equivalence. In 1970s, we started from Galileo's Equivalence Principle and derived its consequences for an electromagnetic system whose Lagrangian density is $L_I^{(EM)}$ + $L_I^{(EM-P)}$ + $L_I^{(P)}$ where $L_I^{(EM)}$ and $L_I^{(EM-P)}$ are defined in (5) & (6) and the particle Lagangian density $L_I^{(P)}$ defined as $-\Sigma_I m_I (ds_I)/(dt) \delta(\mathbf{x}-\mathbf{x}_I)$ with $m_I$ the mass of the *I*th (charged) particle and $s_I$ its 4-line element from the metric $g_{ij}$ [20, 28]. The result is that the constitutive tensor density $\chi^{ijkl}$ can be constrained and expressed in metric form with additional pseudoscalar (axion) field $\varphi$:

$$\chi^{ijkl} = (-g)^{1/2}[(1/2)g^{ik} g^{jl} - (1/2)g^{il} g^{kj}] + \varphi e^{ijkl}, \qquad (7)$$

where $g^{ij}$ is the inverse of $g_{ij}$ and $g$ = det ($g_{ij}$). Thus the particle metric $g_{ij}$ induces the constitutive tensor density $\chi^{ijkl}$ and generates the light cone for electromagnetic wave propagation. We notice that the axion field does not contribute to the ray propagation in the lowest-order eikonal approximation [29-33].

However, there are two aspects of this derivation which are not quite satisfying. First, the constitutive tensor must match the particle metric in a certain way from the Galileo's Equivalence Principle: the metric is not constructed directly from the constitutive tensor. This is not quite satisfying from the theoretical point of view. Second, the Galileo's Equivalence Principle is only verified to high precision for unpolarized test bodies. The tests on polarized bodies which include polarized electromagnetic energy are not very precise. This is not quite satisfying from experimental point of view. Noticing that the pulses from pulsars propagating in the galactic gravitational field arrive at earth at the same time independent of polarization, i.e., are nonbirefringent, we began to use the nonbirefringence condition as a starting point in the later part of 1970's [29-31]. This would be satisfying in two aspects. First, the equal-time arrival of pulses independent of polarization can be formulated as a statement of Galilio's Equivalence Principle for photons and is therefore theoretically satisfying. Second, the nonbirefringence is tested to high precision with pulses from pulsars and is therefore experimentally satisfying. From the non-birefringent propagation in spacetime to high precision, we constrain the constitutive tensor density to core metric form and construct the light cone of electromagnetic wave propagation with an additional scalar (dilaton) field $\psi$ and an additional pseudoscalar (axion) field $\varphi$ [29-31]. The theoretical condition for no birefringence (no splitting, no retardation) for electromagnetic wave propagation in all directions is that the constitutive tensor density $\chi^{ijkl}$ can be written in the following form

$$\chi^{ijkl} = (-h)^{1/2}[(1/2)h^{ik} h^{jl} - (1/2)h^{il} h^{kj}]\psi + \varphi e^{ijkl}, \qquad (8)$$

where $h^{ij}$ is a metric constructed from $\chi^{ijkl}$ ($h$ = det ($h_{ij}$) and $h_{ij}$ the inverse of $h^{ij}$) which generates the light cone for electromagnetic wave propagation [29-32]. We constructed the relation (8) in the weak-violation/weak-field approximation of the Einstein Equivalence Principle (EEP) and applied to pulsar observations in 1981 [29-31]; Haugan and Kauffmann [32] reconstructed the relation (8) and applied to radio galaxy observations in 1995. After the cornerstone work of Lämmerzahl and Hehl [33], Favaro and Bergamin [34] finally proved the relation (8) without assuming weak-field approximation (see also Dahl [35]). Polarization measurements of electromagnetic waves from pulsars and cosmologically distant astrophysical sources yield stringent constraints agreeing with (8) down to $2\times10^{-32}$ fractionally (for a review, see [25, 26]).

The complete agreement with EEP for photon sector requires (i) no birefringence; (ii) no polarization rotation; (iii) no amplification/no attenuation in spacetime propagation. With no birefringence, any skewonless spacetime constitutive tensor must be of the form (8) as we just reviewed in the last paragraph. In the next section, we show that with the condition of no polarization rotation and the condition of no amplification/no attenuation satisfied, the axion $\varphi$ and the dilaton $\psi$ should be constant respectively. That is, no varying axion field and no varying dilaton field respectively; the EEP for photon sector is observed; the spacetime constitutive tensor is of metric-induced form only. Thus we tie the three observational conditions to EEP and to metric-induced spacetime constitutive tensor in the photon sector. This is aesthetic and more satisfying. Previously we have worked out the condition on polarization rotation for axion field [21, 25, 26]. In the next section, we work out the no amplification/no attenuation condition for dilaton field.

To tie the light cone metric to matter metric, we need Hughes-Drever-type experiments and Eötvös-type experiments on matter bodies with various proportions of electromagnetic energy contents and other energy contents [29, 31, 36]. These kinds of experiments have been done to ultra-precision in the laboratory and will also be done to even higher precision in space. Eötvös-type experiments also test the variation of dilaton field. The current upper limit from Eötvös-type experiments on the variation of dilaton field in the earth environment and in the solar system is about $10^{-10}$ in terms of relevant gravitational potential [25].

In section 2, we study the electromagnetic wave propagation in the dilaton field and the axion field with the constitutive tensor given by

$$\chi^{ijkl} = (-g)^{1/2}[(1/2)g^{ik}\,g^{jl} - (1/2)g^{il}\,g^{kj}]\psi + \varphi e^{ijkl}. \tag{9}$$

In section 3, we study the cosmic constraints on the dilaton field from CMB spectrum observations. In section 4, we give a few comments and present an outlook.

## 2. Wave propagation and the dispersion relation in dilaton and axion field

In this section, we derive the electromagnetic wave propagation and the dispersion relation in dilaton and axion field. Let us begin with the general problem of wave propagation in electrodynamics (1a,b) with constitutive relation (2) for explaining and fixing the scheme. The sourceless Maxwell equation (1b) is equivalent to the local existence of a 4-potential $A_i$ such that

$$F_{ij} = A_{j,i} - A_{i,j}, \tag{10}$$

with a gauge transformation freedom of adding an arbitrary gradient of a scalar function to $A_i$. The Maxwell equation (1a) in vacuum with (3) is then

$$(\chi^{ijkl}A_{k,l})_{,j} = 0. \tag{11}$$

Using the derivation rule, we have

$$\chi^{ijkl}A_{k,l,j} + \chi^{ijkl}{}_{,j} A_{k,l} = 0. \tag{12}$$

(i) For slowly varying, nearly homogeneous field/medium, and/or (ii) in the eikonal approximation with typical wavelength much smaller than the gradient scale and time-variation scale of the field/medium, the second term in (12) can be neglected compared to the first term, and we have

$$\chi^{ijkl}A_{k,lj} = 0. \tag{13}$$

This approximation is usually called the eikonal approximation. In this approximation, the dispersion relation is given by the generalized covariant quartic Fresnel equation (see, e.g. [9]). It is well-known that axion does not contribute to this dispersion relation [9, 25, 26, 29-33]. Dilaton does not contribute to this dispersion relation either. The generalized Fresnel equation is algebraic and homogeneous in the wave covector. Since the dilaton only gives a multiplicative scalar factor in the equation, it does not change the dispersion relation.

To derive the influence of the dilaton field and the axion field on the dispersion relation, one needs to keep the second term in equation (12). This has been done for the axion field in references [21, 25, 26, 37, 38, 39]. Here we develop it for the joint dilaton field and axion field. Near the origin in a local inertial frame, the constitutive tensor density in dilaton field $\psi$ and axion field $\varphi$ [equation (9)] becomes

$$\chi^{ijkl}(x^m) = [(1/2)\, \eta^{ik}\, \eta^{jl} - (1/2)\, \eta^{il}\, \eta^{kj}]\, \psi(x^m) + \varphi(x^m)\, e^{ijkl} + \mathrm{O}(\delta_{ij}x^ix^j), \tag{14}$$

where $\eta^{ij}$ is the Minkowski metric with signature $-2$ and $\delta_{ij}$ the Kronecker delta. In the local inertial frame, we use the Minkowski metric and its inverse to raise and lower indices. Substituting (14) into the equation (12) and multiplying by 2, we have

$$\psi\, A^{i,j}{}_j + \psi\, A^{j,i}{}_j + \psi_{,j}\, A^{i,j} - \psi_{,j}\, A^{j,i} + 2\, \varphi_{,j}\, e^{ijkl}\, A_{k,l} = 0. \tag{15}$$

We notice that (15) is both Lorentz covariant and gauge invariant.

We expand the dilaton field $\psi(x^m)$ and the axion field $\varphi(x^m)$ at the 4-point (event) $P$ with respect to the event (time and position) $P_0$ at the origin as follows:

$$\psi(x^m) = \psi(P_0) + \psi_{,i}(P_0)\, x^i + \mathrm{O}(\delta_{ij}x^ix^j), \tag{16a}$$

$$\varphi(x^m) = \varphi(P_0) + \varphi_{,i}(P_0)\, x^i + \mathrm{O}(\delta_{ij}x^ix^j). \tag{16b}$$

To look for wave solutions, we use eikonal approximation which does not neglect field gradient/medium inhomogeneity. We choose $z$-axis in the wave propagation direction so that the solution takes the following form:

$$A \equiv (A_0, A_1, A_2, A_3) = (\underline{A}_0, \underline{A}_1, \underline{A}_2, \underline{A}_3)\, \mathrm{e}^{ikz-i\omega t} = \underline{A}_i\, \mathrm{e}^{ikz-i\omega t}. \tag{17}$$

We expand the solution as

$$A_i = A^{(0)}{}_i + A^{(1)}{}_i + O(2) = [\underline{A}^{(0)}{}_i + \underline{A}^{(1)}{}_i + O(2)]\, e^{ikz-i\omega t} = \underline{A}_i\, e^{ikz-i\omega t}. \tag{18}$$

Now use eikonal approximation to obtain a local dispersion relation. In the eikonal approximation, we only keep terms linear in the derivative of the dilaton field and the axion field; we neglect terms containing the second-order derivatives of the dilaton field or the axion field, terms of $O(\delta_{ij} x^i x^j)$ and terms of mixed second order, e.g. terms of $O(A^{(1)}{}_i x^j)$ or $O(A^{(1)}{}_I \psi_{,j})$; we call all these terms $O(2)$.

Imposing radiation gauge condition in the zeroth order in the weak field/dilute medium approximation, we find to zeroth order, (15) is

$$\psi A^{(0)ij}{}_{,j} = 0, \text{ or } A^{(0)ij}{}_{,j} = 0, \tag{19}$$

and the corresponding zeroth order solution and the dispersion relation are

$$A^{(0)}{}_i = (0, A^{(0)}{}_1, A^{(0)}{}_2, 0) = \underline{A}^{(0)}{}_i\, e^{ikz-i\omega t} = (0, \underline{A}^{(0)}{}_1, \underline{A}^{(0)}{}_2, 0)\, e^{ikz-i\omega t}, \tag{20a}$$

$$\omega = k + O(1). \tag{20b}$$

Substituting (19) and (20a,b) into equation (15), we have

$$\psi A^{(0)ij}{}_{,j} + \psi A^{(1)ij}{}_{,j} + \psi A^{(1)ji}{}_{,j} + \psi_{,j} A^{(0)ij} - \psi_{,j} A^{(0)ji} + 2\varphi_{,j} e^{ijkl} A^{(0)}{}_{k,l} = 0 + O(2). \tag{21}$$

The $i = 0$ and $i = 3$ components of (21) give

$$\psi A^{(1)0j}{}_{,j} + \psi A^{(1)j0}{}_{,j} + \psi_{,j} A^{(0)0,j} - \psi_{,j} A^{(0)j,0} + 2\varphi_{,j} e^{0jkl} A^{(0)}{}_{k,l} = 0 + O(2), \tag{22a}$$

$$\psi A^{(1)3j}{}_{,j} + \psi A^{(1)j3}{}_{,j} + \psi_{,j} A^{(0)3,j} - \psi_{,j} A^{(0)j,3} + 2\varphi_{,j} e^{3jkl} A^{(0)}{}_{k,l} = 0 + O(2). \tag{22b}$$

Noticing that $A^{(0)0} = A^{(0)3} = 0$ from (20) and that $A^{(1)0j}{}_{,j} = -(\omega^2 - k^2) A^{(1)0} = O(1) \times A^{(1)0} = O(2)$, we have

$$A^{(1)j}{}_{,j}{}^0 = \psi^{-1} (\psi_{,j} A^{(0)j,0} - 2\varphi_{,j} e^{0jkl} A^{(0)}{}_{k,l}) + O(2), \tag{23a}$$

$$A^{(1)j}{}_{,j}{}^3 = \psi^{-1} (\psi_{,j} A^{(0)j,3} - 2\varphi_{,j} e^{3jkl} A^{(0)}{}_{k,l}) + O(2). \tag{23b}$$

Noting that taking the derivative with respect to the upper index 0 (upper index 3) is to multiply by $-i\omega$ ($ik$), we simplify (23a) and (23b); they both give to the same equation

$$A^{(1)j}{}_{,j} = -\psi^{-1}(\psi_{,1} - 2\varphi_{,2}) A^{(0)}{}_1 - \psi^{-1}(\psi_{,2} + 2\varphi_{,1}) A^{(0)}{}_2 + O(2). \tag{24}$$

Since equation (24) does not contain $\omega$ and $k$, it does not contribute to the determination of the dispersion relation. We regard equation (24) as the modified Lorentz gauge condition in the O(1) order in the dilaton field and the axion field.

Using the gauge condition (24), we obtain the $i = 1$ and $i = 2$ components of equation (21) as

$$(\omega^2 - k^2)\underline{A}^{(0)}{}_1 - i k \underline{A}^{(0)}{}_1 \psi^{-1}(\psi_{,0} + \psi_{,3}) - 2 i k \underline{A}^{(0)}{}_2 \psi^{-1}(\varphi_{,0} + \varphi_{,3}) = 0 + \text{O}(2), \quad (25a)$$
$$(\omega^2 - k^2)\underline{A}^{(0)}{}_2 - i k \underline{A}^{(0)}{}_2 \psi^{-1}(\psi_{,0} + \psi_{,3}) + 2 i k \underline{A}^{(0)}{}_1 \psi^{-1}(\varphi_{,0} + \varphi_{,3}) = 0 + \text{O}(2). \quad (25b)$$

These two equations determine the dispersion relation. To have a nontrivial solution, we must have the following determinant vanishes:

$$\det \begin{bmatrix} (\omega^2 - k^2) - i k \psi^{-1}(\psi_{,0} + \psi_{,3}) & -2 i k \psi^{-1}(\varphi_{,0} + \varphi_{,3}) \\ 2 i k \psi^{-1}(\varphi_{,0} + \varphi_{,3}) & (\omega^2 - k^2) - i k \psi^{-1}(\psi_{,0} + \psi_{,3}) \end{bmatrix}$$
$$= [(\omega^2 - k^2) - i k \psi^{-1}(\psi_{,0} + \psi_{,3})]^2 - 4 k^2 \psi^{-2}(\varphi_{,0} + \varphi_{,3})^2 = 0 + \text{O}(2). \quad (26)$$

Equation (26) is the dispersion relation in the dilaton field and the axion field. Its solutions in $\omega^2$ are

$$\omega^2 = k^2 - i k \psi^{-1}(\psi_{,0} + \psi_{,3}) \pm 2 k \psi^{-1}(\varphi_{,0} + \varphi_{,3}) + \text{O}(2). \quad (27)$$

The solutions in $\omega$ or $k$ are

$$\omega = k - (i/2)\psi^{-1}(\psi_{,0} + \psi_{,3}) \pm \psi^{-1}(\varphi_{,0} + \varphi_{,3}) + \text{O}(2), \quad \text{or} \quad (28a)$$
$$k = \omega + (i/2)\psi^{-1}(\psi_{,0} + \psi_{,3}) \pm \psi^{-1}(\varphi_{,0} + \varphi_{,3}) + \text{O}(2). \quad (28b)$$

The group velocity is

$$v_g = \partial\omega/\partial k = 1, \quad (29)$$

independent of polarization. When the dispersion relation is satisfied, (25a) and (25b) have two independent solutions for the polarization eigenvectors $\underline{A}^{(0)}{}_i = (\underline{A}^{(0)}{}_0, \underline{A}^{(0)}{}_1, \underline{A}^{(0)}{}_2, \underline{A}^{(0)}{}_3)$ with

$$\underline{A}^{(0)}{}_0 = 0; \quad (30a)$$
$$\underline{A}^{(0)}{}_1/\underline{A}^{(0)}{}_2 = [2 i k \psi^{-1}(\varphi_{,0} + \varphi_{,3})] / [(\omega^2 - k^2) - i k \psi^{-1}(\psi_{,0} + \psi_{,3})]$$
$$= [2 i k \psi^{-1}(\varphi_{,0} + \varphi_{,3})] / [\pm 2 k \psi^{-1}(\varphi_{,0} + \varphi_{,3})] = \pm i; \quad (30b)$$
$$\underline{A}^{(0)}{}_0 = 0, \quad (30c)$$

for $\omega = k - (i/2) \psi^{-1} (\psi_{,0} + \psi_{,3}) \pm \psi^{-1} (\varphi_{,0} + \varphi_{,3}) + O(2)$ respectively. From (30b), the two polarization eigenstates are left circularly polarized state and right circularly polarized state in varying axion field. This agrees with the electromagnetic wave propagation in axion field as derived earlier [21, 25, 26, 37, 38, 39].

With the dispersion (28), the plane-wave solution (17) propagating in the z-direction is

$$A \equiv (A_0, A_1, A_2, A_3) = (0, \underline{A}^{(0)}{}_1, \underline{A}^{(0)}{}_2, 0)\, e^{ikz-i\omega t}$$
$$= (0, \underline{A}^{(0)}{}_1, \underline{A}^{(0)}{}_2, 0) \exp[ikz - ikt \pm (-i)\, \psi^{-1} (\varphi_{,0}\, t + \varphi_{,3}\, z) - (1/2)\, \psi^{-1} (\psi_{,0}\, t + \psi_{,3}\, z)], \quad (31)$$

with $\underline{A}^{(0)}{}_1 = \pm i\, \underline{A}^{(0)}{}_2$. The additional factor acquired in the propagation is $\exp[\pm (-i)\, \psi^{-1} (\varphi_{,0}\, t + \varphi_{,3}\, z)] \times \exp[-(1/2)\psi^{-1} (\psi_{,0}\, t + \psi_{,3}\, z)]$. The first part of this factor, i.e., the axion factor $\exp[\pm (-i)\, \psi^{-1} (\varphi_{,0}\, t + \varphi_{,3}\, z)]$ adds a phase in the propagation. The second part of this factor, i.e., the dilaton factor $\exp[- (1/2)\, \psi^{-1} (\psi_{,0}\, t + \psi_{,3}\, z)]$ amplifies or attenuates the wave according to whether $(\psi_{,0}\, t + \psi_{,3}\, z)$ is less than zero or greater than zero. For the right circularly polarized electromagnetic wave, the effect of the axion field in the propagation from a point $P_1 = \{x_{(1)}{}^i\} = \{x_{(1)}{}^0; x_{(1)}{}^\mu\} = \{x_{(1)}{}^0, x_{(1)}{}^1, x_{(1)}{}^2, x_{(1)}{}^3\}$ to another point $P_2 = \{x_{(2)}{}^i\} = \{x_{(2)}{}^0; x_{(2)}{}^\mu\} = \{x_{(2)}{}^0, x_{(2)}{}^1, x_{(2)}{}^2, x_{(2)}{}^3\}$ is to add a phase of $\alpha = \psi^{-1} [\varphi(P_2) - \varphi(P_1)]$ ($\approx \varphi(P_2) - \varphi(P_1)$ for $\psi \approx 1$) to the wave; for left circularly polarized light, the effect is to add an opposite phase (Ni 1973). Linearly polarized electromagnetic wave is a superposition of circularly polarized waves. Its polarization vector will then rotate by an angle $\alpha$. The effect of the dilaton field is to amplify with a factor $\exp[- (1/2)\, \psi^{-1} (\psi_{,0}\, t + \psi_{,3}\, z)] = \exp[- (1/2) ((\ln \psi)_{,0}\, t + (\ln \psi)_{,3}\, z)] = (\psi(P_1)/\psi(P_2))^{1/2}$. The dilaton field contributes to the amplitude of the propagating wave is positive or negative depending on $\psi(P_1)/\psi(P_2) > 1$ or $\psi(P_1)/\psi(P_2) < 1$ respectively.

For plane wave propagating in direction $n^\mu = (n^1, n^2, n^3)$ with $(n^1)^2 + (n^2)^2 + (n^3)^2 = 1$, the solution is

$$A(n^\mu) \equiv (A_0, A_1, A_2, A_3) = (0, \underline{A}_1, \underline{A}_2, \underline{A}_3) \exp(-i\, kn^\mu x_\mu - i\omega t)$$
$$= (0, \underline{A}_1, \underline{A}_2, \underline{A}_3) \exp[-ikn^\mu x_\mu - ikt \pm (-i)\psi^{-1}(\varphi_{,0}\, t - n^\mu \varphi_{,\mu} n_\nu x^\nu) - (1/2)\, \psi^{-1}(\psi_{,0} t + n^\mu \psi_{,\mu} n_\nu x^\nu)], \quad (32)$$

where $\underline{A}_\mu = \underline{A}^{(0)}{}_\mu + n_\mu n^\nu \underline{A}^{(0)}{}_\nu$ with $\underline{A}^{(0)}{}_1 = \pm i\, \underline{A}^{(0)}{}_2$ and $\underline{A}^{(0)}{}_3 = 0$ as in (31).

The above analysis is local. In the global situation, choose local inertial frames along the wave trajectory and integrate along the trajectory. Since $\psi$ is a scalar, the integration gives $(\psi(P_1)/\psi(P_2))^{1/2}$ as the amplification factor for the propagation in the dilaton field. For small dilaton field variations, the amplification/attenuation factor is equal to [1 − (Δψ/ψ)] to a very good approximation with $\Delta\psi \equiv \psi(P_2) - \psi(P_1)$. Since this factor does not depend on the wave number/frequency and polarization, it will not distort the source spectrum in propagation, but gives an overall amplification/attenuation factor to the spectrum. The axion field contributes to the phase factor and induces polarization rotation as in previous investigations [21, 25, 26, 37, 38]. For $\psi \approx 1$ (constant), the induced polarization rotation agrees with previous results which were obtained without considering dilaton effect. If the dilaton field varies significantly, a $\psi$-weight needs to be included in the integration.

### 3. Constraints on cosmic dilaton field

In this section we look into the observations/experiments to constrain the dilaton field contribution to spacetime constitutive tensor density. From equation (31) and (32) in the last section, we have derived that the amplitude and phase factor of propagation in the cosmic dilaton and cosmic axion field is changed by

$$(\psi(P_1) / \psi(P_2))^{1/2} \exp[ikz - ikt \pm (-i)(\varphi(P_1) - \varphi(P_2))t].$$

The effect of dilaton field is to give amplification ($\psi(P_1) - \psi(P_2) > 0$) or attenuation ($\psi(P_1) - \psi(P_2) < 0$) to the amplitude of the wave independent of frequency and polarization.

The spectrum of the cosmic microwave background (CMB) is well understood to be Planck blackbody spectrum. In the cosmic propagation, this spectrum would be amplified or attenuated by the factor $(\psi(P_1) / \psi(P_2))^{1/2}$. However, the CMB spectrum is measured to agree with the ideal Planck spectrum at temperature $2.7255 \pm 0.0006$ K [40] with a fractional accuracy of $2 \times 10^{-4}$. The spectrum is also red-shifted due to cosmological curvature (or expansion), but this does not change the blackbody character. The measured shape of the CMB spectra does not deviate from Planck spectrum within its experimental accuracy. In the dilaton field the relative increase in power is proportional to the amplitude increase squared, i.e., $\psi(P_1)/\psi(P_2)$. Since the total power of the blackbody radiation is proportional to the temperature to the fourth power $T^4$, the fractional change of the dilaton field since the last scattering surface of the CMB must be less than about $8 \times 10^{-4}$ and we have

$$|\Delta\psi|/\psi \leq 4\,(0.0006/2.7255) \approx 8 \times 10^{-4}. \tag{33}$$

Direct fitting to the CMB data with the addition of the scale factor $\psi(P_1)/\psi(P_2)$ would give a more accurate value.

## 4. Discussion and outlook

*In section 3, we conclude that the dilaton field is constrained by the agreement and accurate calibration of cosmic microwave background (CMB) to blackbody radiation so that the fractional change $|\Delta\psi|/\psi$ of dilaton since the last CMB scattering surface is less than $8 \times 10^{-4}$.* This completes our study for the EEP for the photon sector in the $\chi$-framework (constitutive tensor framework) in the skewonless case. In the following, we highlight a few points and discuss various issues to be studied further.

(i) Just like axionic degree of freedom in spacetime/medium gives polarization rotation in the electromagnetic wave propagation independent of frequency (energy) and the state of linear polarization, the dilatonic degree of freedom gives amplification/attenuation in the electromagnetic wave propagation independent of frequency (energy) and polarization.

(ii) In the constitutive tensor density approach to spacetime structure of electromagnetism, from experiments/observations for the photon sector there are three steps to construct the core metric and to constrain the remaining scalar/pseudoscalar degrees of freedom empirically: (a) from the nonbirefringence in electromagnetic wave propagation, we reach the core metric plus dilaton plus axion structure; (b) from the no amplification/attenuation in the electromagnetic wave propagation independent of frequency (energy) and polarization, the dilaton degree of freedom can be constrained; and (c) from the no polarization rotation independent of frequency (energy) and the state of linear polarization in the electromagnetic wave propagation, the axionic degree of freedom can be constrained. To the extent these three steps are successful, Einstein Equivalence Principle would be verified or new physics would be discovered.

(iii) In (ii), (a) is the fundamental step. It does not depend on metrology and standard. It only depends on coincidence. This is the most important step. In (ii), (c) needs a local inertial frame as reference frame to measure the angle of polarization rotation [23]. The constraints on CPR angle $\alpha$ and axion field $\Delta\varphi$ from the UV polarization observations of radio galaxies are about a couple of degree (0.03 rad). Those from the CMB polarization observations [23-26] are about 0.02 for CPR mean value $|\langle\alpha\rangle|$ and 0.03 for the CPR fluctuations $\langle(\alpha - \langle\alpha\rangle)^2\rangle^{1/2}$. More precise constraints/measurements from PLANCK Surveyor are expected soon. In (ii), (b) needs metrology system and standards to establish the amplitude measurement accuracy. This is guaranteed by the Universality of Metrology and Standards which is an alternate statement of the Einstein Equivalence Principle [41]. The experimental constraint of (b) gives empirical supports to the Universality of Metrology and Standards in the cosmic scale. To the extent the dilaton field varies, the metrology depends on position and time as noted by Dicke [42]. Nontrivial dilaton field may also related to the possibility of the variability of vacuum impedance or the variability of the fine structure constant [43].

(iv) In the laboratory, dilaton field is constrained by the ultra-precision Eötvös-type and Galileo-type WEP experiments. The fractional variation of dilaton field is constrained to less than about $10^{-10}$ fractionally. Space experiments will improve on the precision. This compliments the cosmic constraint on dilaton field.

(v) In the very early universe, dilaton is sometimes postulated to explain the inflation. The implication of these inflation models to the subsequent evolution of dilaton field to the last scattering surface and thereafter should be thoroughly investigated as it could be assessable to experimental tests

(vi) In view of the existence of axionic material [27], searches for dilatonic material in macroscopic electrodynamics would also be interesting and would foster mutual interactions in the two fields. This study may also find application in macroscopic electrodynamics in the case inhomogeneous media which contain a dilatonic degree and an axionic degree of freedom in their constitutive tensor density.

(vii) For slowly varying, nearly homogeneous field/medium, and/or in the lowest eikonal approximation with typical wavelength much smaller than the gradient scale and time-variation scale of the field/medium, the second term in (12) can be neglected compared to the first term. For inhomogeneous field/medium, and/or in the higher eikonal approximation, the second term in (12) cannot be neglected. A thorough study of the dispersion generalizing the method used here would be warranted.

(viii) Since gravity is universal, in previous papers [44, 45], we have explored the foundations of classical electrodynamics using $\chi$-framework. In this paper, we have looked more into the dilaton field in comic propagation. This contributes to improved cosmological tests of classical electrodynamics.

**Acknowledgements**

We would like to thank Yu. N. Obukhov, and F. W. Hehl for helpful comments on the manuscript. We would also like to thank the National Science Council (Grant No. NSC102-2112-M-007-019) and the National Center for Theoretical Sciences (NCTS) for supporting this work in part.

**References**

[1] H. Minkowski, Konig. Ges. Wiss. Gottingen. Math.-Phys. (1908) 53-111.

[2] A. Einstein, Jahrb. Radioakt. 4 (1907) 411-462.

[3] A. Einstein, Eine Neue Formale Deutung der Maxwellschen Feldgleichungen der Elektrodynamik, *Königlich Preußische Akademie der Wissenschaften* (Berlin), 184-188 (1916); See also, A new formal interpretation of Maxwell's field equations of Electrodynamics, in *The Collected Papers of Albert Einstein*, Vol. 6, A. J. Kox et al., eds. (Princeton University Press: Princeton, 1996) pp. 263–269.

[4] See also F. W. Hehl, Ann. Phys. (Berlin) **17**, No. 9-10, 691-704 (2008) for a historical account and detailed explanation.

[5] H. Weyl, Raum-Zeit-Materie, Springer, Berlin, 1918; See also, Space-Time-Matter, English translation of the 4th German edition of 1922 (Dover, Mineola, New York, 1952).

[6] F. Murnaghan, The absolute significance of Maxwell's eruations, *Phys. Rev.* 17, 73-88 (1921).

[7] F. Kottler, Maxwell'sche Gleichungen und Metrik. *Sitzungsber. Akad. Wien IIa* **131**, 119-146 (1922),.

[8] É Cartan, Sur les variétés à connexion affine et la Théorie de la Relativité Généralisée, *Annales scientifirues de l'École Normale Supérieure* **40** pp. 325-412, **41** pp. 1-25, **42** pp. 17-88 (1923/1925); See also, On Manifolds with an Affine Connection and the Theory of General Relativity, English translation of the 1955 French edition. Bibliopolis, Napoli, 1986.

[9] F. W. Hehl and Yu. N. Obukhov, *Foundations of Classical Electrodynamics: Charge, Flux, and Metric* (Birkhäuser: Boston, MA, 2003).

[10] F. W. Hehl, Yu. N. Obukhov, G. F. Rubilar, On a possible new type of a T-odd skewon field linked to electromagnetism, in: A. Macias, F. Uribe, E. Diaz (Eds.), Developments in Mathematical and Experimental Physics, Volume A: Cosmology and Gravitation (Kluwer Academic/Plenum, New York, 2002) pp. 241-256 [gr-rc/0203096].

[11] F. W. Hehl, Yu. N. Obukhov, G. F. Rubilar, Int. J. Mod. Phys. A 17 (2002) 2695.

[12] Yu. N. Obukhov and F. W. Hehl, Phys. Rev. D 70 (2004) 125015.

[13] F. W. Hehl and Yu. N. Obukhov, Phys. Lett. A 334 (2005) 249-259.

[14] F. W. Hehl, Yu. N. Obukhov, G. F. Rubilar and M. Blagojevic, Phys. Lett. A 347 (2005) 14.

[15] Y.Itin, J. Phys. A: Math. Theor. 42 (2009) 475402.

[16] Y. Itin, Phys. Rev. D 88 (2013) 107502.

[17] W.-T. Ni, Phys. Lett. A (2014) 1217-1223.

[18] Y. Itin, Electromagnetic media with Higgs-type spontaneously broken transparency, arXiv:1406.3442v1.

[19] Y. Itin, On skewon modification of light cone structure, arXiv:1407.6722v1.

[20] W.-T. Ni, *Phys. Rev. Lett.* **38**, 301, (1977).

[21] W.-T. Ni, A Nonmetric Theory of Gravity, preprint, Montana State University (1973) [http://astrod.wikispaces.com/].

[22] W.-T. Ni, Spin, Torsion and Polarized Test-Body Experiments, in *Proceedings of the 1983 International School and Symposium on Precision Measurement and Gravity Experiment*, Taipei, Republic of China, January 24-February 2, 1983, ed. by W.-T. Ni (Published by National Tsing Hua University, Hsinchu, Taiwan, Republic of China) pp. 531-540 [http://astrod.wikispaces.com/].

[23] S. di Serego Alighieri, W.-T. Ni, W.-P. Pan, Astrophys. J., 792 (2014) 35.

[24] S. di Serego Alighieri, Cosmological Birefringence: an Astrophysical test of Fundamental Physics, Proceeding of Symposium I of JENAM 2010 – Joint European and National Astronomy Meeting: From Varying Couplings to Fundamental Physics, Lisbon, 6-10 Sept. 2010, Editors C. Martins and P. Molaro (Springer-Verlag, Berlin, 2011) p.139 [arXiv:1011.4865].

[25] W.-T. Ni, *Reports on Progress in Physics* 73 (2010) 056901.

[26] W.-T. Ni, *Prog. Theor. Phys. Suppl.* **172** (2008) 49 [arXiv:0712.4082].

[27] F. W. Hehl, Yu. N. Obukhov, J.-P. Rivera and H. Schmid, Phys. Rev. A 77 (2008) 022106.

[28] W.-T. Ni, *Bull. Am. Phys. Soc.* **19**, 655 (1974).

[29] W.-T. Ni, Equivalence Principles and Precision Experiments, in *Precision Measurement and Fundamental Constants II*, ed. by B. N. Taylor and W. D. Phillips, Natl. Bur. Stand. (U S) Spec. Publ. 617 (1984) pp. 647-651.

[30] W.-T. Ni, Timing Observations of the Pulsar Propagations in the Galactic Gravitational Field as Precision Tests of the Einstein Equivalence Principle, in *Proceedings of the Second Asian-Pacific Regional Meeting of the International Astronomical Union on Astronomy, Bandung, Indonesia – 24 to29 August 1981*, ed. by B. Hidayat and M. W. Feast (Published by Tira Pustaka, Jakarta, Indonesia, 1984) pp. 441-448.

[31] W.-T. Ni, Equivalence Principles, Their Empirical Foundations, and the Role of Precision Experiments to Test Them, in *Proceedings of the 1983 International School and Symposium on Precision Measurement and Gravity Experiment*, Taipei, Republic of China, January 24-February 2, 1983, ed. by W.-T. Ni (Published by National Tsing Hua University, Hsinchu, Taiwan, Republic of China, 1983) pp. 491-517 [http://astrod.wikispaces.com/].

[32] M. Haugan and T. Kauffmann, Phys. Rev. D 52 (1995) 3168.

[33] C. Lämmerzahl and F. W. Hehl, *Phys. Rev. D* 70 (2004) 105022.

[34] A. Favaro and L. Bergamin, Annalen der Physik 523 (2011) 383-401.

[35] M. F. Dahl, Journal of Physics A: Mathematical and Theoretical 45 (2012)

405203.

[36] W.-T. Ni, Implications of Hughes-Drever Experiments, in *Proceedings of the 1983 International School and Symposium on Precision Measurement and Gravity Experiment*, Taipei, Republic of China, January 24-February 2, 1983, ed. by W.-T. Ni (Published by National Tsing Hua University, Hsinchu, Taiwan, Republic of China, 1983) pp. 519-529 [http://astrod.wikispaces.com/].

[37] W.-T. Ni, Chin. Phys. Lett. 22 (2005) 33-35.

[38] Y. N. Obukhov, F. W. Hehl, Phys. Lett. A 341 (2005) 357.

[39] Y. Itin, Gen. Rel. Grav. 40 (2008) 1219.

[40] D. J. Fixsen, Astrophys. J. 707 (2009) 916.

[41] W.-T. Ni, Some Basic Points About Metrology, in *Proceedings of the 1983 International School and Symposium on Precision Measurement and Gravity Experiment*, Taipei, Republic of China, January 24-February 2, 1983, ed. by W.-T. Ni (Published by National Tsing Hua University, Hsinchu, Taiwan, Republic of China, 1983) pp. 121-134 [http://astrod.wikispaces.com/].

[42] R. H. Dicke, The theoretical significance of experimental relativity (Gordon and Breach: New York, NY, 1964)

[43] F.W. Hehl, Y. N. Obukhov, Gen. Rel. Grav. 37 (2005) 733

[44] W.-T. Ni, Foundations of Electromagnetism, Equivalence Principles and Cosmic Interactions, Chaper 3 in *Trends in Electromagnetism - From Fundamentals to Applications*, pp. 45-68 (March, 2012), Victor Barsan and Radu P. Lungu (Ed.), ISBN: 978-953-51-0267-0, InTech (open access) (2012) [arXiv:1109.5501], Available from:

http://www.intechopen.com/books/trends-in-electromagnetism-from-fundamentals-to-applications/foundations-of-electromagnetism-eruivalence-principles-and-cosmic-interactions.

[45] W.-T. Ni, H.-H. Mei and S.-J. Wu, Mod. Phys. Lett. 28 (2013) 1340013.